\documentclass[10pt,final,journal,letterpaper,twoside,twocolumn]{IEEEtran}

\usepackage{cite}
\usepackage{graphicx,color,epsfig,rotating}
\usepackage{amsfonts,amsmath,amssymb,bbm}
\usepackage{algorithm, algorithmic}
\usepackage{subfigure}
\usepackage{amsmath}
\usepackage{cite}
\usepackage{mdwtab}
\usepackage{subfigure}
\usepackage{placeins}
\usepackage{psfrag, graphicx}
\usepackage[latin1]{inputenc}
\usepackage{amssymb}
\usepackage{makeidx}
\usepackage{url}

\long\def\comment#1{}

\newfont{\bbb}{msbm10 scaled 700}

\newfont{\bb}{msbm10 scaled 1100}

\newcommand{\EE}{\mbox{\bb E}}

\newcommand{\Fc}{{\cal F}}

\newcommand{\Hc}{{\cal H}}

\newcommand{\Kc}{{\cal K}}

\newcommand{\Nc}{{\cal N}}

\newcommand{\Rc}{{\cal R}}

\newcommand{\Tc}{{\cal T}}
\newcommand{\Uc}{{\cal U}}

\newcommand{\muv}{\hbox{\boldmath$\mu$}}

\newcommand{\be}{\begin{equation}}
\newcommand{\ee}{\end{equation}}
\newcommand{\bea}{\begin{eqnarray}}
\newcommand{\eea}{\end{eqnarray}}

\newtheorem{defn}{Definition}
\newtheorem{theorem}{Theorem}

\begin{document}

\title{Caching Eliminates the Wireless Bottleneck in Video-Aware Wireless Networks}

\author{ Andreas F. Molisch,$^{1}$ Giuseppe Caire,$^{1}$ David Ott,$^{2}$ Jeffrey R. Foerster,$^{2}$ Dilip Bethanabhotla,$^{1}$ Mingyue Ji,$^{1}$ 

\thanks{$^{1}$ Department of Electrical Engineering, University of Southern California, Los Angeles, CA; 

${2}$ Intel Corporate Research
}

}

\maketitle

\thispagestyle{empty}
\pagestyle{empty}

\vspace{-0.5cm}

\begin{abstract}

Cellular data traffic almost doubles every year, greatly straining network capacity. The main driver for this development is wireless video. Traditional methods for capacity increase (like using more spectrum and increasing base station density) are very costly, and do not exploit the unique features of video, in particular a high degree of
{\em asynchronous content reuse}. In this paper we give an overview of our work that proposed and detailed a new transmission paradigm exploiting content reuse, and the fact that storage is the fastest-increasing quantity in modern hardware. Our network structure uses caching in helper stations (femto-caching) and/or devices, combined with highly spectrally efficient
short-range communications to deliver video files. For femto-caching, we develop optimum storage schemes and dynamic 
streaming policies that optimize video quality. For caching on devices, combined with device-to-device communications, we show that 
communications within {\em clusters} of mobile stations should be used; the cluster size can be adjusted to optimize the tradeoff between
frequency reuse and the probability that a device finds a desired file cached by another device in the same cluster. We show that in many situations the network throughput increases linearly with the number of users, and that D2D communications also is superior in providing a better tradeoff between throughput and outage than traditional base-station centric systems. Simulation results with realistic numbers of users and channel conditions show that 
network throughput (possibly with outage constraints) can be increased by two orders of magnitude compared to conventional schemes. 

\end{abstract}




\section{Introduction}

\begin{figure}[ht]
\centerline{\includegraphics[width=7cm]{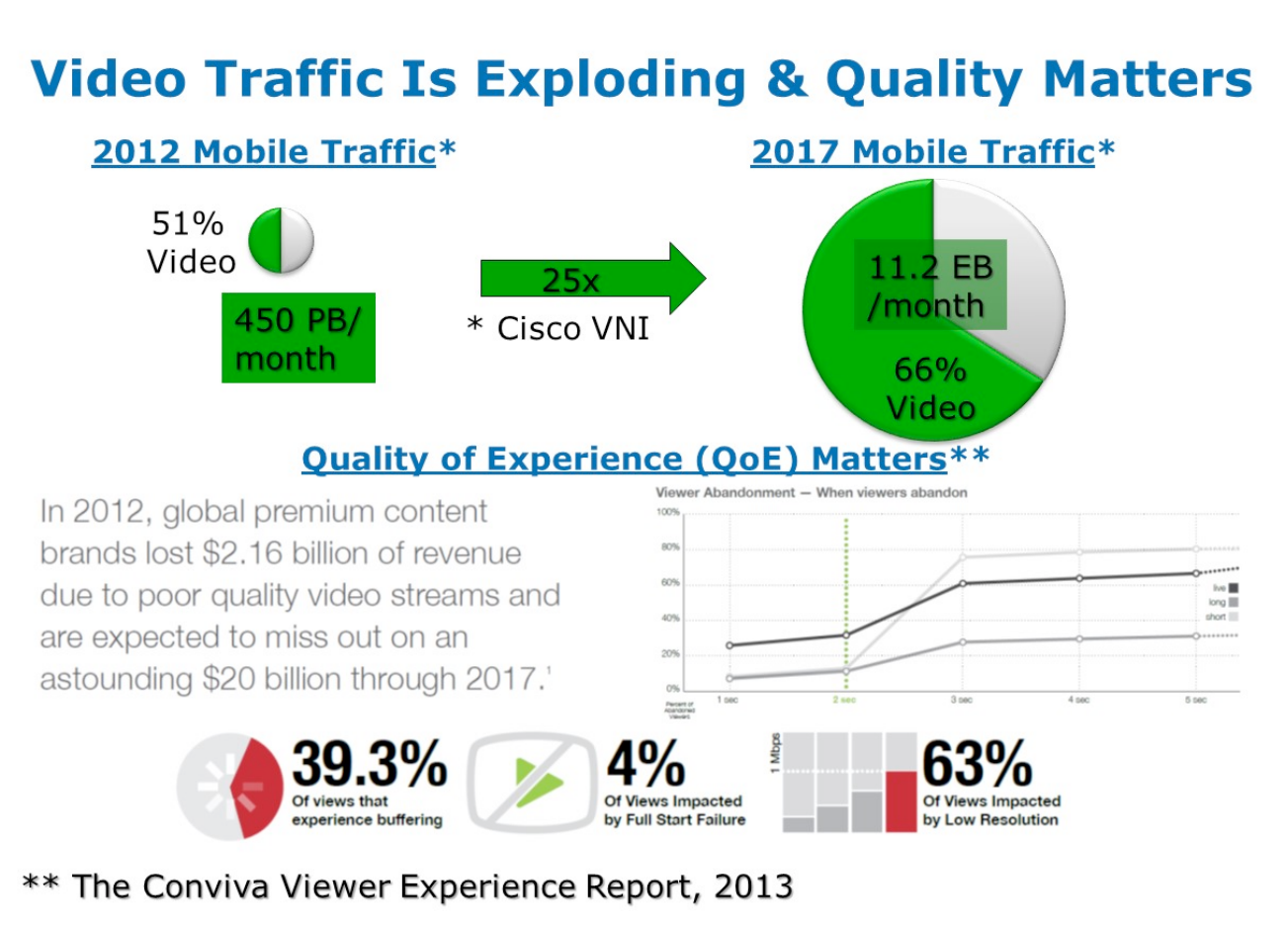}}
\caption{Demand for video traffic will continue to grow significantly and QoE has clear financial implications according to industry sources Cisco \cite{Cisco1} and Conviva \cite{Conviva}}
\label{fig: Fig1}
\end{figure}

Demand for video content over wireless networks has grown significantly in recent years and shows no sign of letting up.  According to the {\em Cisco Visual Networking Index} mobile forecast for 2012-2017 \cite{Cisco1}, mobile video data is expected to grow at a compound annual growth rate of 75 percent to 7.4 exabyes (one million gigabytes) by 2017.  By this time, it is expected to be 66.5 percent of global mobile traffic data (11.2 exabytes), up from 51 percent in 2012 (see Fig. \ref{fig: Fig1}).  We expect both broadcast and on-demand services will continue to expand, including traditional services like streaming TV content (e.g., sporting events) and newer services like video Twitter, video blogging, cloud-based live video broadcasting, and mobile-to-mobile video conferencing and sharing.  Meanwhile, hardware platforms (smart phones, tablets, notebooks, television/set-top boxes, in-vehicle infotainment systems) continue to push the envelope in performance and graphical quality.  More capable processors, better performing graphics, increased storage capacities, and larger displays make devices more powerful and intelligent than ever before.  And with this increase in device capability comes a corresponding increase in demand for high-quality video data; for example, increasing demand for high-definition (HD) and 3D data types.

The implications of these trends for future wireless networks are significant.  While continued evolution in spectral efficiency is to be expected, the maturity of MIMO, air interfaces using OFDM/OFMDA, and Shannon capacity-approaching codes mean that such spectral efficiency improvements will not deliver the increased capacity needed to support future demand for video data.  Additional measures like the brute force expansion of wireless infrastructure (number of cells) and the licensing of more spectrum, while clearly addressing the problem of network capacity, may be prohibitively expensive, require significant time to implement, or be infeasible due to prior spectrum allocations which are not easily modified.

Recognizing these challenges, Intel and several industry partners jointly developed a program to explore non-incremental, systems-level solutions through university research.  Known as {\em Video-aware Wireless Networks} or simply VAWN, the program considers various approaches to enabling a higher capacity in future wireless networks, and in enabling a higher quality of user experience for video and video-based services delivered over wireless networks to intelligent mobile devices. Broad strategies explored in the program include unconventional optimizations in video transport within the network, optimizations in video processing to reduce network transmission requirements and improve user experience, and novel network architectures better suited to address future capacity and quality of service challenges specific to video. 

The approach taken by the group at the University of Southern California (including several of the authors), exploits a unique feature 
of wireless video, namely the high degree of (asynchronous) content reuse. Based on the fact that storage is 
cheap and ubiquitous in today's wireless devices, this group developed a new network structure that is based on {\em replacing backhaul by caching}. This approach, first proposed in  \cite{VAWN_2010}, and expounded and refined in a series of papers \cite{golrezaei2011wireless, golrezaei2012device, golrezaei2013femtocaching, Golrezaei_et_al_2011Infocom, golrezaei2012wireless, Shanmugan_et_al_2013_IT, kim2013adaptive, bethanabhotla2013joint, bethanabhotla2014joint, golrezaei2012device_Globecom,  golrezaei2012base, Mingyue-D2D-ITW, ji2013throughput, JiCaireMolisch2013}), is at the center of the present overview. 

A first approach for exploiting asynchronous content reuse, termed {\em Femto-Caching}, uses dedicated ``helper nodes'' that can cache popular files 
and serve requests from wireless users by enabling localized wireless communication. Such helper nodes
are similar to femto-BSs, but they have two key differences: they have {\em added} a large storage,\footnote{ Note that storage space has
become exceedingly cheap: 2 TByte of data storage capacity,
enough to store 1000 movies, cost only about \$100.} while they {\em do not have or need} a high-speed backhaul. An even higher density of caching can be achieved by using devices themselves as video caches - in other words, using devices such as tablets and laptops (which nowadays have ample storage) as mobile helper stations \cite{golrezaei2012device}. 
The simplest way of using this storage would have each user cache the most popular files. However, this approach is not efficient because many users are interested in similar files, and thus the same videos will be duplicated
on a large number of devices. On the other hand, the cache on each device is too small to cache a reasonably large number of files. 
Thus, it is preferable that the devices "pool" their caching resources, so that different devices cache different files, and then exchange them, when the occasion 
arises, through short-range, highly spectrally efficient, device-to-device (D2D) communications. 
If a requesting device does not find the file in its neighborhood (or in its own cache), 
it obtains the file in the traditional manner from the base station (the base station can also control any occuring D2D communications). 

The remainder of the paper is organized as follows: in Section II, we describe video coding and video streaming techniques, as well as content reuse and viewing habits. The principle of the new network structure is described in Sec. III. The placement of files in helper nodes and devices is discussed in Sec. IV. Fundamental results about throughput and outage in networks with helper stations and D2D communications are described in Secs. V and VI, respectively. Conclusions in Sec. VII round off the paper.

\section{Dynamically Managing Video Quality of Experience}

\subsection {Video Streaming and Quality Management}
Wireless channels are inherently dynamic and time varying depending on a number of factors: (i) movement of device (walking, driving), (ii) changes in the reflectors in the environment (people moving, objects moving), (iii) changes in location (insider, outside), (iv) changes in selected wireless network (WiFi, cellular), and (v) changes in the amount of traffic using the network (i.e., congestion).  For data and web-based applications, some latency due to changes in available network capacity, while annoying, can be tolerated.  However, for video-based applications (especially interactive video conferencing,  but also - depending on buffering capability - for video playback), simply treating data communications as latency tolerant is not sufficient.  In order to maintain an acceptable quality of experience (QoE), it is necessary to adapt the rate of the streamed video using techniques that take into account such factors as the type of video being streamed (fast motion, complex scenes, interactive), the available capacity of the network, time variations in network and channel state, client device information (screen size, etc.) and playback buffer state.  This section will describe some mechanisms for achieving this dynamic adaptation and the role of emerging standards.

Fig. \ref{fig: Fig2} below shows a simplified view of an end-to-end system, including a video server on the left, an end rendering device on the right, and a network lying in between. (Note that video streaming applications are the focus here.)  Labels are included that identify potential opportunities for managing video traffic in intelligent ways. To accommodate different devices and to support multiple streaming rates, multiple copies (formats, bitrates) of the video content is stored on the server.  Alternatively, the video can be transcoded on the fly.  The decision of whether to transcode or store multiple copies depends on cost, complexity, and performance tradeoffs, and must take into account the facts of memory and compute resources in the underlying system.  It may also depend on the popularity of the content and where the content is stored within the data center or network. 

The availability of multiple video streaming rates makes possible dynamic adaptation during a streaming session in response to changes in wireless channel state. Today, multiple copies of the same video provide a range of bitrates to a client device which can choose among them.  To improve user playback experience, however, as well as to improve the efficiency of data storage and transport, we believe QoE will be important in the future.  Measures of QoE may take into account the quality of the displayed video (resolution, compression artifacts), re-buffering events, and lost packets.  QoE metrics provide an alternative to throughput-based approaches which rely on the often mistaken assumption that higher bitrates mean higher quality.  A key challenge here, however, is effectively estimating video quality independent of bitrate.  Fortunately, a great deal of progress has been made recently by researchers estimating video quality based on both device and content characteristics (see \cite{Foerster2013, choivideo, liao2013achieving}).  This creates new opportunities for optimizing the end-to-end system when tighter coordination between the video server, network, and end devices can be realized. 

Enhancements to emerging standards are helping to promote QoE-based optimization within end-to-end systems.  In particular, standards supporting Dynamic Adaptive Streaming over HTTP (DASH) are being developed by the MPEG and 3GPP standards bodies (see \cite{ISO1, 3GPP1, ISO2, ISO3, ISO4, Sodagar2011, stockhammer2011dynamic, oyman2012quality, Ramamurthi2014, Oyman2013, Oyman20132, Oyman20133}).  Two recent additions to these standards are (1) the inclusion of QoE feedback metrics from the device to the network, and (2) support for providing QoE metrics along with video content that is sent to a device.  (In some cases, video QoE metrics can also be computed directly by the end device.)  These additions are important because they enable better system-wide optimization of video transport based on the end user QoE.  For example, the device can decide which future segments to request based on the current status of its playback buffer and known quality levels of upcoming segments.  This supports a more intelligent balancing of playback quality and re-buffering risk.  The network can also make more informed decisions on how to allocate available bandwidth across multiple competing video flows by optimizing the quality jointly across all of them.  Using rate-distortion information (a measure of video quality) and playback buffer state for each flow, for instance, a network scheduler can implement QoE-based resource allocation as an alternative to standard proportionally fair throughput schemes.

\begin{figure}[ht]
\centerline{\includegraphics[width=7cm]{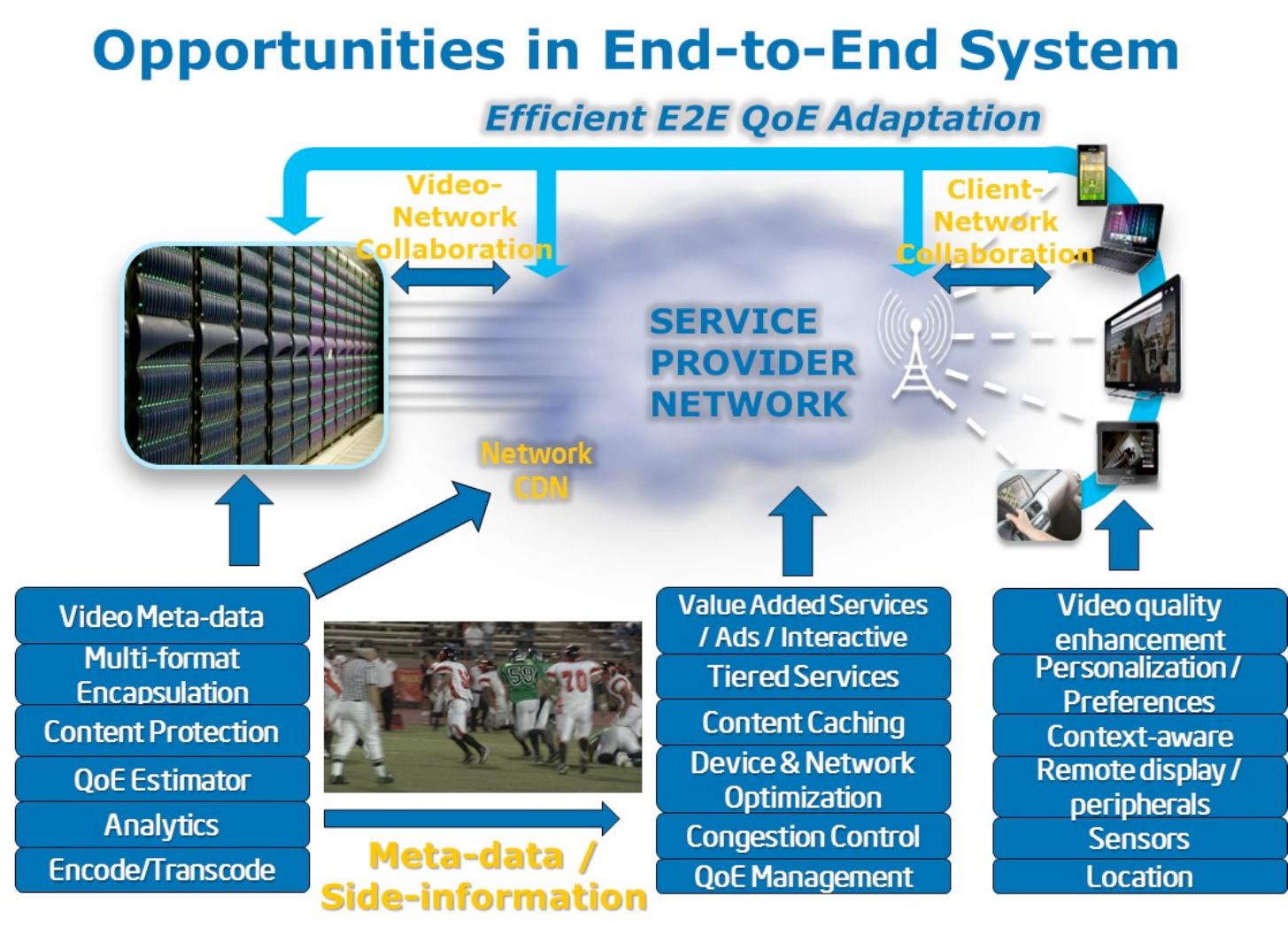}}
\caption{Simplified end-to-end system for video streaming. }
\label{fig: Fig2}
\end{figure}
%

\subsection{Content Reuse}

Wireless Video distinguishes itself from other wireless content through its strong content reuse, i.e., the same content is seen 
by a large number of peoples. However, in contrast to TV, the bulk of wireless video traffic is due to asynchronous {\em video on demand}, where users request video files from some
cloud-based server at arbitrary times. As indicated in Sec. I, 
the use of caching enables to exploit {\em content overlap}, even in the presence {\em asynchronism of requests}.
In other words, a few popular videos (YouTube clips, sports highlights, and movies) 
account for a considerable percentage of video traffic on the Internet, even though they are viewed at different times by different people.  
Numerous experimental studies have indicated that Zipf distributions are good models for the measured popularity of video files \cite{zipf,tracedata}. Under this model, the frequency of the $i$-th popular file, denoted by $P_r(f)$, is inversely proportional to its rank:
 \begin{equation}\label{zipf}
 P_r(f)=\frac{{\frac{1}{{{f^{\gamma_r} }}}}}{{\sum\limits_{j = 1}^m {\frac{1}{{{j^{\gamma_r} }}}} }},\,\,\ 1\leq f \leq m.
 \end{equation}
 The Zipf exponent $\gamma_r$ characterizes the distribution by controlling the relative popularity of files.
 Larger $\gamma_r$ exponents correspond to higher content reuse, \textit{i.e.,} the first few popular files account for the majority of requests.
 Here, $m$ is the size of the library of files that are of interest to the set of considered users (note that the library size can be a function of the number of considered users $n$; we assume in the following $m$ increase like $n^{\alpha}$, where $\alpha\ge 0$). 
  
A further important property of the library is that it changes only on a fairly slow timescale (several days or weeks); it can furthermore be shaped by content providers, e.g., through pricing policies, or other means. 

Note, however, some caveats concerning the general applicability of the work in the remainder of the paper. 
It applies principally to a setting where a content library of relatively large files (e.g., movies and TV shows) is refreshed 
relatively slowly (e.g., on a daily basis), and where the number of users consuming such a library is significantly larger than the number of items in the library.
This may apply to a possible future implementation of movie services, while collections of short videos (like YouTube) show wider ranges of interests. 
In short, this paper reflects a set of results and approaches that are relevant in the case where the caching phase (placement of content in the caches) occurs with a clear time-scale separation with respect to the delivery phase (the process of delivering video packets for streaming to the users), 
and where the size of the content library is moderate with respect to the users' population.

\section{Network Structure}

\subsection{Helper Stations and File Requests}

\begin{figure}
\centering
\includegraphics[height = 50mm]{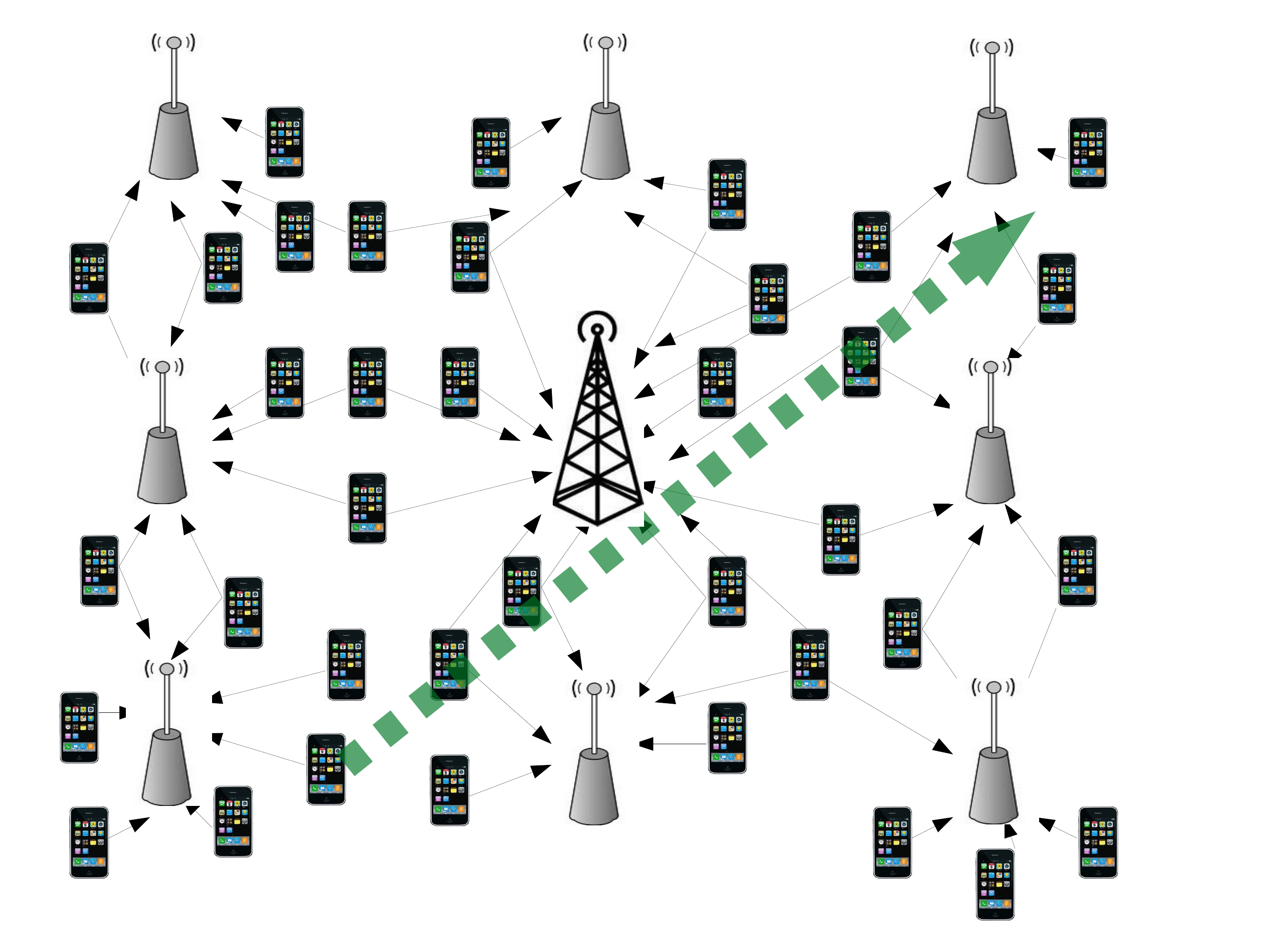}
\caption{A sample network scenario with 64 helpers and 320 users and a mobile user (green path).}
\label{topology}
\end{figure}

We first consider the network structure with helper stations. The wireless network consists of  multiple helper stations $\Hc$, talking to multiple users $\Uc$; a central base station may be present to serve users that cannot find the files they want in the helper stations. An example network is shown in Fig.~\ref{topology}. Each user requests a video file from a library $\Fc$ of possible files. We denote the set of helpers in the vicinity of user $u$ as $\Nc(u)$. Similarly, $\Nc(h)$ denotes the set of users in the vicinity of helper $h$. The helpers may not have access to the whole video library, because of backhaul constraints or caching constraints. In general, we denote by $\Hc(f)$ the set of helpers that contain file $f \in \Fc$. Hence, user $u$ requesting file $f_u$ can only download video chunks from helpers in the set $\Nc(u) \cap \Hc(f_u)$. In Section~\ref{dash-section-dilip}, we consider the problem of devising a dynamic scheduling scheme such that helpers feed the video files sequentially (chunk by chunk) to the requesting users. Given the high density of helpers, any user is typically in the range of multiple helpers. Hence, in order to cope with user-helper association, load balancing and inter-cell interference, an efficient video streaming policy is described in Section~\ref{dash-section-dilip} which allows the users to dynamically select the helper node to download from, and determine adaptively the video quality level of the download.

\subsection{Device-to-Device (D2D) Caching Networks}

\begin{figure}
\centering
\includegraphics[width = 7cm]{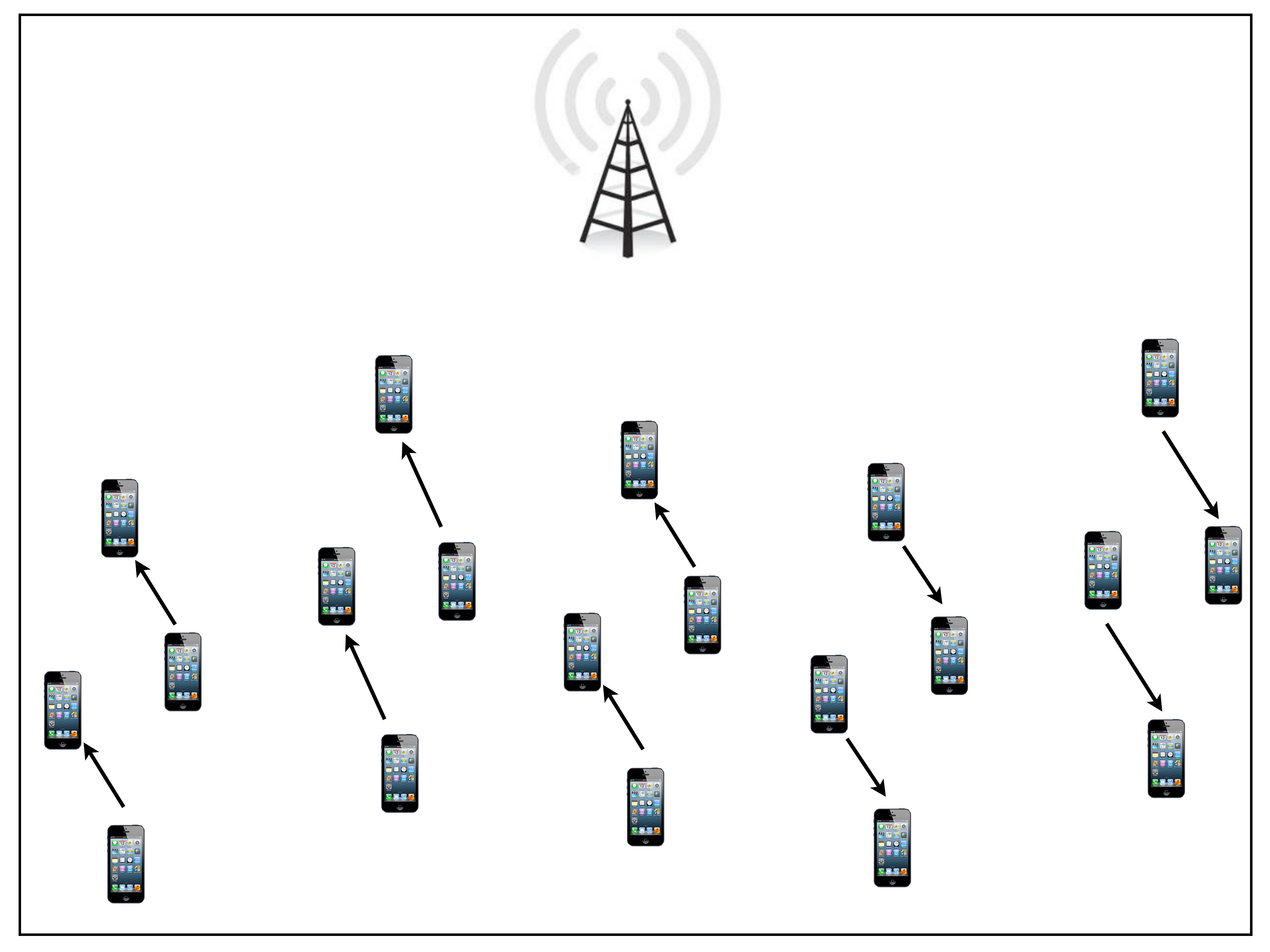}
\caption{An illustration of D2D caching networks, where each user (device) can cache $M$ files and we let users be served through high-spectral-efficiency D2D links.}
\label{fig: D2D_Network_overview}
\end{figure}

\begin{figure}
\centering
\subfigure[]{
\centering \includegraphics[width=3.5cm]{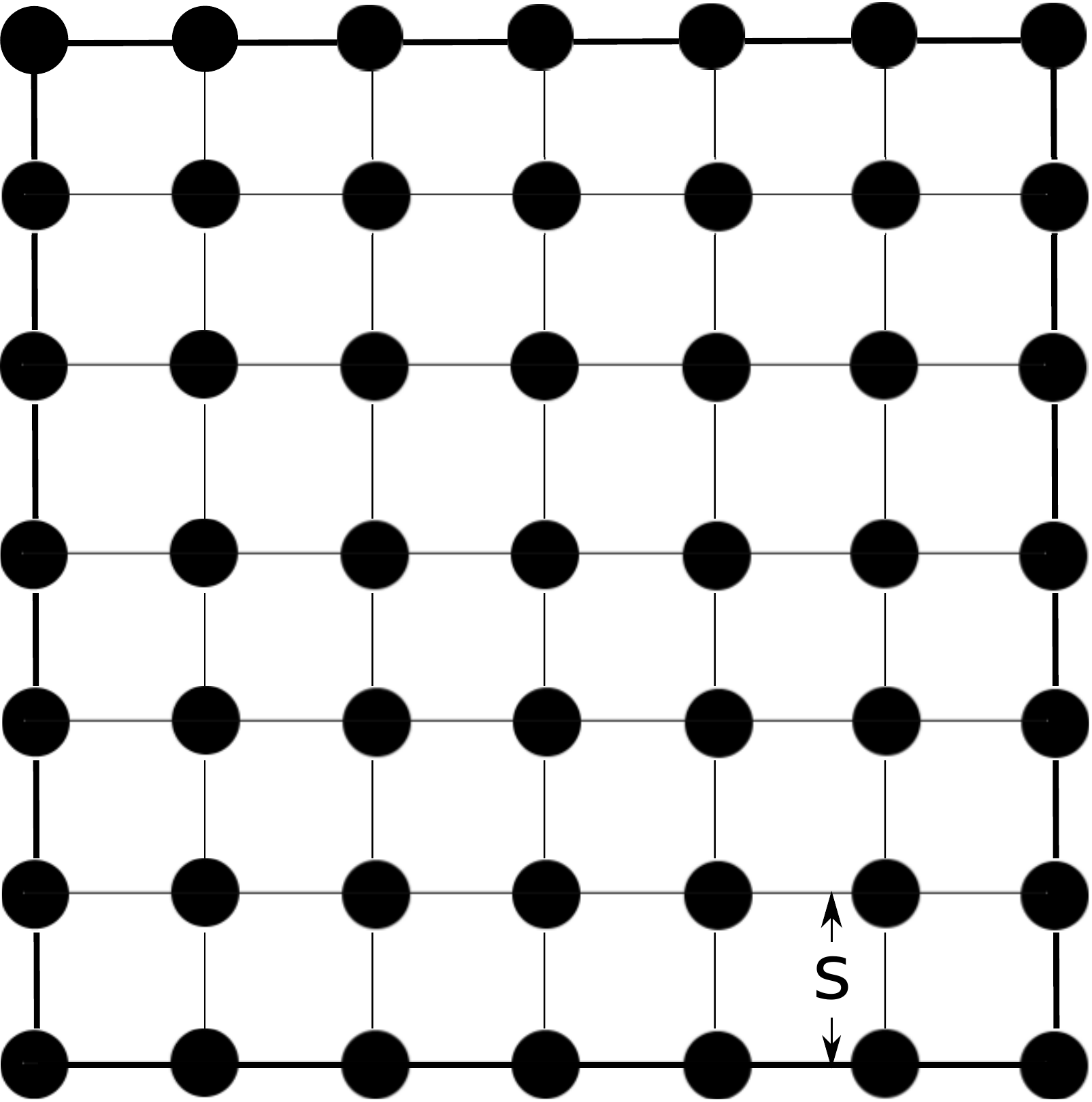}
\label{fig: Grid_Network_D2D}
}
\subfigure[]{
\centering \includegraphics[width=3.5cm]{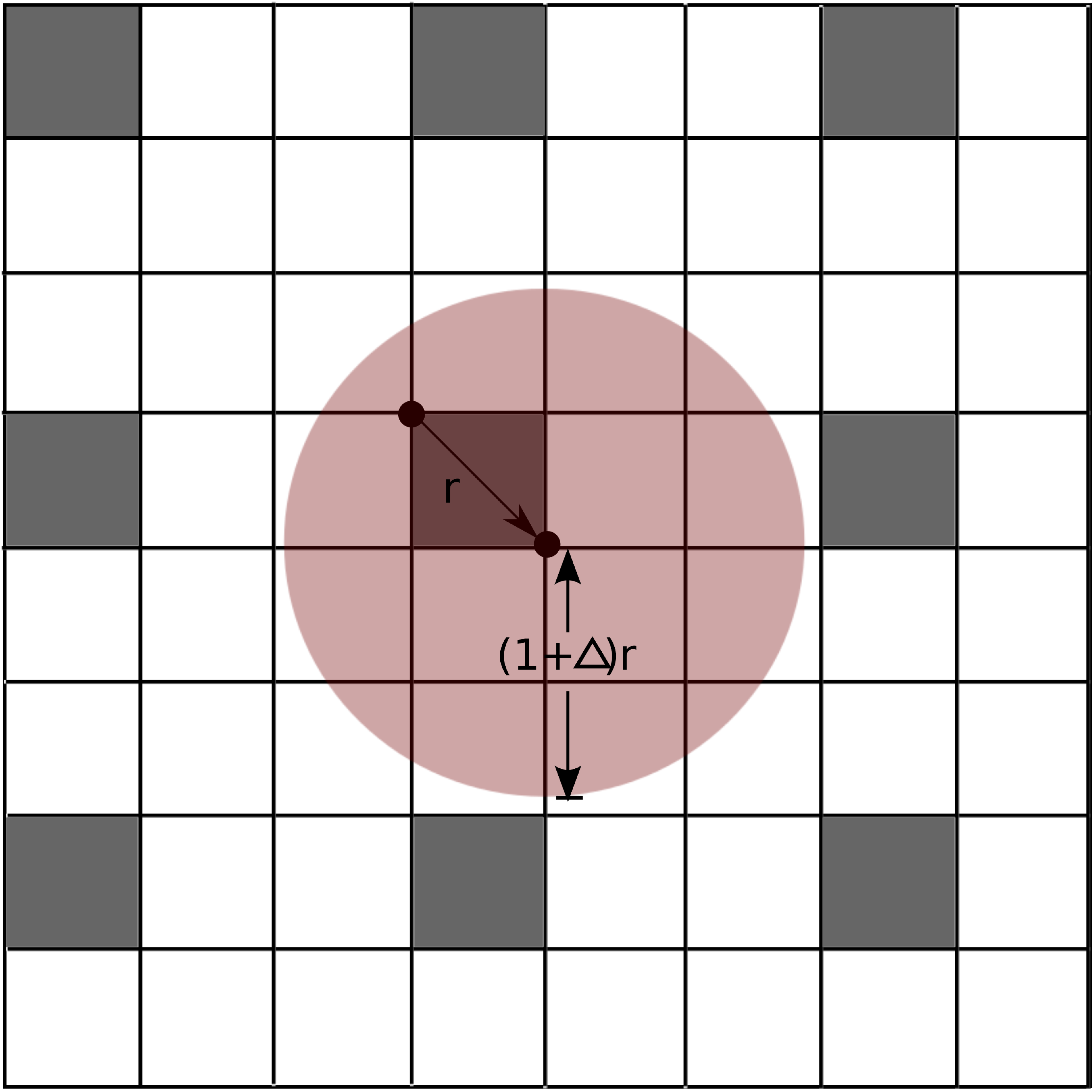}
\label{fig: Grid_TDMA}
}
\caption{a)~Grid network with $n=49$ nodes (black circles) with minimum separation $s = \frac{1}{\sqrt{n}}$. 
b)~An example of single-cell layout and the interference avoidance TDMA scheme. 
In this figure, each square represents a cluster. 
The gray squares represent the concurrent transmitting clusters. 
The red area is the disk where the protocol model imposes no other concurrent transmission. 
$r$ is the worst case transmission range and $\Delta$ is the interference parameter. 
We assume a common $r$ for all the transmitter-receiver pairs. 
In this particular example, the TDMA parameter is $K=9$.}
\end{figure}

When users also have the ability of prefetching (video) files, instead of requesting the files from the base station or the helpers, we allow users make requests from other users and get served via high-spectral-efficiency D2D links (see Fig. \ref{fig: D2D_Network_overview}). If the D2D links are not available for some users (see Section \ref{sec: Theoretical Scaling Laws analysis}), then these unserved users are treated as in outage and in practice, they can be simply served by the base station or the helpers. To make the network model tractable, we consider the transmission of the video files instead of streaming, and neglect the issue of rate adaptation. In addition, we consider a simple gird structure, which is formed by $n$ user nodes $\Uc = \{1, \ldots, n\}$ placed on a regular grid on the unit square, with minimum distance $1/\sqrt{n}$. (see Fig.~\ref{fig: Grid_Network_D2D}; we will replace this grid structure by the uniform distribution of the nodes when mentioned specifically.).
Let each user $u \in \Uc$ request a file $f \in \Fc = \{1, \ldots, m\}$ in an i.i.d. manner, 
according to a given request probability mass function $P_r(f)$, which is assumed to be a Zipf distribution given by (\ref{zipf}) with 
parameter $0 < \gamma_r < 1$ \cite{breslau1999web}. 
Moreover, we let each user cache $M$ files. 
The BS keeps track of which devices can communicate with each other, and which files are cached on each device. Such BS-controlled D2D communication is more efficient (and more acceptable to spectrum owners if the communications occur in a licensed band) than traditional uncoordinated peer-to-peer communications.

Communications between nodes follow the protocol model \cite{gupta2000capacity}:\footnote{ In the simulations of Section \ref{sec: Simulation Results}, we relax the protocol model constraint and take interference into consideration by treating it like noise.} namely, 
transmission between user nodes $u$ and $v$ is possible if their distance $d(u,v)$ is less than or equal to some fixed transmission range $r$, 
and if there is no other active transmitter within distance $(1 + \Delta) r$ from destination $v$, where 
$\Delta > 0$ is the interference control parameter. 
Successful transmissions can take place at rate $C_r$ bit/s/Hz, which is a non-increasing function of the transmission range $r$ \cite{Shanmugan_et_al_2013_IT}. 
In this model, we do not consider power control (which would allow different transmit powers, and thus transmission ranges), for each user. 
Moreover, we treat $r$ as a design parameter that can be set as a function of $m$ and $n$.\footnote{Since the number of possibly requested files $m$ typically varies with the number of users in the system $n$, and $r$ can vary with $n$, $r$ can also be a function of $m$.} All communications are assumed to be single-hop (see also Section \ref{sec: Performance of D2D Caching Networks}). These model assumptions allow for a sharp analytical characterization of the throughput scaling law including the leading constants. In Section \ref{sec: Performance of D2D Caching Networks}, we will see that the schemes designed by this simple model yields
promising performance also in realistic channel propagation and interference conditions. 

For many of our derivations, we furthermore subdivide the cell into equal-sized, disjoint groups of users that we call "clusters" of size (radius) $r$, with $g_c$ nodes in it. To further simplify the mathematical model, we assume that only nodes that are part of the same cluster can communicate with each other.  If a user can find the requested file inside the cluster, we say there is one \emph{potential link} in this cluster; when at least one link is scheduled, we say that the cluster is "active". We use an {\em interference avoidance} scheme, such that at most one link can be active in each cluster on one time-frequency resource.

\section{File placement}

The proposed system operates in two steps: (i) file placement (caching) and (ii) delivery. These two processes happen on different timescales: the cache content needs to change only on a timescale of days, weeks, or months, i.e., much slower than the actual delivery to the users. Thus, caches could be filled either through a very slow backhaul, or through cellular connection at night time, when the spectral resources are not required for other purposes. 

\subsection{File placement in helper stations}

\begin{figure}
\includegraphics[width=8cm]{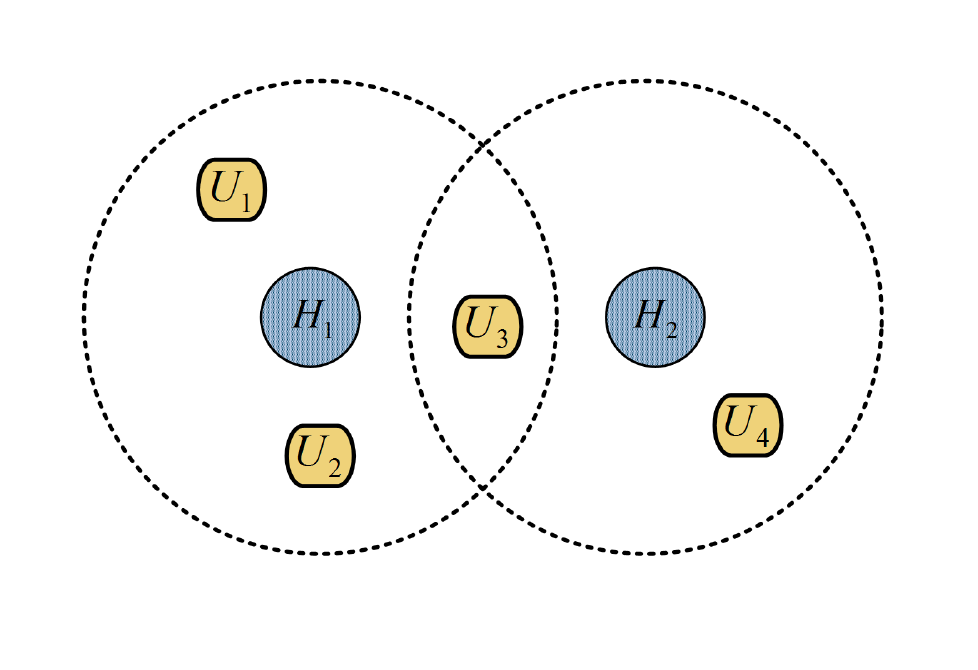}
\caption{Example for caching when users have access to multiple helpers.}
\label{fg_DistCaching}
\end{figure}

We start out with the case where complete files are stored in the helper stations. If the distance between helpers is large, and each MS can connect only to a single helper, each helper should cache the most popular files, in sequence of popularity, until its cache is full. However, when each MS can communicate with multiple helpers, the question on how to best assign files to different helpers becomes a more complicated. Consider the case in Figure \ref{fg_DistCaching}. Users $U_1$ and $U_2$ would prefer helper $H_1$ to cache the $M$ most popular files since this minimizes their expected downloading time. Similarly, user $U_4$ would prefer that helper $H_2$ also caches the $M$ most popular files. However $U_3$ would prefer $H_1$ to cache the $M$ most popular files and $H_2$ the second $M$ most popular (or the opposite), thus creating a distributed cache of size $2M$ for user $U_3$. Thus we can see that in the distributed caching problem, the individual objectives of different users may be in conflict, and we need sophisticated algorithms to find an optimum assignment.  

Let us assume for the moment that 1) the network topology is known; 2) the long-term average link rates are known; 3) the user demand distribution (file popularity) is known. However, the actual demands are not known beforehand, so that caching placement must be done only based on the {\em statistics} of the user requests. Our goal is to minimize the average download time. 
We distinguish further between uncoded and coded caching In the uncoded case, video-encoded files are cached directly (with the possibility of storing the same file in multiple locations).  
In the coded case, we consider placing coded chunks of the files on different helper stations, such that obtaining any sufficiently large number of these chunks allows reconstruction of the original video file (e.g., using the scheme in \cite{DRESS}).

In \cite{Shanmugan_et_al_2013_IT} we showed that the uncoded-placement problem is NP-complete. However, it can be formulated as the maximization of a monotone submodular function over matroid constraints, for which  a simple 
greedy strategy achieves at least $\frac{1}{2}$ of the optimum value.  For the coded case, the optimum cache placement can be formulated as a convex optimization problem, for which optimum solutions can be found through efficient algorithms. In general, the optimum value of delay obtained with the coded optimization is better than the uncoded optimization 
because any placement matrix with integer entries is a feasible solution to the coded problem. 
In this sense, the coded optimization is a convex relaxation of the  uncoded problem.

We conclude this section by mentioning that the conditions under which we derived the optimum caching are rarely fulfilled in practice. 
While the user demand distribution $P_r(f)$ may be well estimated and predicted, the network topology is typically time-varying with dynamics comparable or faster than the file transmission, therefore reconfiguring the caches at this time scale is definitely not practical. However, further computer experiments have also shown that the cache distribution obtained when the mobile stations are in "typical" distances from the helpers also provides good performances for various other realizations of random placement of nodes.
Furthermore, distributed random caching turns out to be ``good enough'' as we shall see in Sec. VI. 
Hence, comparing optimal placement with random caching
yields useful insight on the potential performance gap lost by a decentralized approach. Interestingly, in any reasonable network 
configuration it turns out that such a gap is very small.

\subsection{File placement for D2D communications}
\label{sec: File placement for D2D communications}

Also for D2D communications, the question of which files should be cached by which user are essential.  Building on the protocol model explained in Sec. III.B, a critical question for each user is whether the file it is interested in can be found within the communication radius $r$ from its current location. In other words, in order to enable D2D communication it is not sufficient that the distance between two users be less than $r$; users should also find their desired files in the cache of another device with which they can communicate. The decision of what to store can be taken in a centralized or distributed way, called deterministic and random 

In deterministic caching a central control (typically the BS) orders the devices to cache specific files. Similar to the situation in femtocaching, we assume that the location of the caching nodes, and the demand distribution, is known. Finding the optimal deterministic file assignment for the general case follows the same principles as for femtocaching outlined above. A simplification occurs when the devices are grouped into {\em clusters} such that only communication within the cluster is possible (for more details see Sec. VI). In this case the deterministic caching algorithm is greatly simplified: the devices in the cluster should simply cache the most popular files in a disjoint manner, i.e., no file should be cached twice in the cluster. Deterministic caching is only feasible if the location of the nodes and the Channel State Information (CSI) is known a priori, and remains constant between the filling of the cache and the actual file transmission; thus it applies only if the caching nodes are fixed wireless devices. It is also useful for providing upper performance bounds for other caching strategies. 
In random caching, each device randomly and independently caches a set of files according to a common probability mass function. In our earlier papers, we assumed that the caching distribution is also a Zipf distribution, though with a parameter $\gamma_c$ that is different from $\gamma_r$, and which has to be optimized for a particular $\gamma_r$ and $r$. Since the Zipf distribution is characterized by a single parameter, this description gives important intuitive insights about how concentrated the caching distribution should be. 

In \cite{ji2013throughput}, we found that the optimal caching distribution $P_c^*$ that maximizes the probability that any user finds its requested file inside 
its own cluster is given (for a node arrangement on a rectangular grid as described above) by
\begin{equation}
\label{eq: optimal caching distribution}
P_c^*(f) = \left[1 - \frac{\nu}{z_{f}}\right]^+,  \;\;\; f = 1,\ldots, m,
\end{equation}
where $\nu = \frac{m^*-1}{\sum_{f=1}^{m^*} \frac{1}{z_{f}}}$, 
$z_{f} = P_r(f)^{\frac{1}{M(g_c - 1)-1}}$, $m^* = \Theta \left (\min \{\frac{M}{\gamma_r}g_c, m\}\right )$ and $[\Lambda]^+ = \max[\Lambda,0]$.

\section{Adaptive Streaming from Helper Stations}\label{dash-section-dilip}
 
We now turn to the delivery phase, in particular for the femtocaching (helper station). We furthermore concentrate on the case that the video files are {\em streamed}, i.e., that replay at the receiver starts before the complete file has been transmitted. Such streaming is widely used for standard video-on-demand systems, using protocols such as Microsoft Smooth Streaming (Silverlight), Apple HTTP Live Streaming, and 3GPP Dynamic Adaptive Streaming over HTTP (DASH). We have adapted such on-demand streaming to our caching architectures, in particular the network setup with helper stations. Dividing each video stream into chunks, we solve the problem of "which user should get a video "chunk", at what quality, from which helper station".  
 
\subsection{Problem formulation}

We represent a video file as a sequence of chunks of equal duration. Each chunk may contain a 
different number of source-encoded bits, due to variable bit-rate (VBR) coding (see Sec. II), and the same video file is encoded at different quality levels, such that lower quality levels correspond to fewer encoded bits. These quantities can vary across video files, and even for the same video they can vary across both chunks and quality levels. For example, the same compression level may produce a different user quality index as well as a different bit requirement from one chunk to the next, depending on if the video chunk is showing a constant blue sky or a busy city street.

 In our system, the requested chunks are queued at the helpers, and each helper $h$ maintains a queue $Q_{hu}$ pointing at each of the users $u$ in its vicinity. We pose the {\em Network Utility Maximization}(NUM) problem of maximizing a concave and component wise non-decreasing network utility function $\phi_u(\cdot)$ of the users' long-term average quality indices $\overline{D}_u$, subject to stability of the queues $Q_{hu}$ at all the helpers. The concavity of the network utility function imposes some desired notion of fairness between the users. The problem formulation is given as:
 \begin{align}
\textrm{maximize}  & \;\;\; \sum_{u}\phi_u(\overline{D}_u) \nonumber \\
 \textrm{subject to} & \;\;\; \overline{Q}_{hu} < \infty~\forall~ (h,u) \label{NUM}
\end{align}
 We solve this problem in \cite{bethanabhotla2013joint} using the Lyapunov Drift Plus Penalty approach and obtain a policy that decomposes naturally into two distinct operations that can be implemented in a decentralized fashion: 1) Congestion control; 2) Transmission scheduling. 
 
 \subsection{Congestion Control} \label{cong-control}
Congestion control decisions are made at each streaming user, which decides from which helper to request the next chunk and at which quality index this shall be downloaded. For every time slot $t$, each $u \in \Uc$ chooses the helper in its neighborhood having the shortest queue, i.e., 
\begin{equation*} \label{helper-selection}
h^*_u(t)  = \mbox{argmin} \left \{  Q_{hu}(t) \; : \;  h \in \Nc(u)\cap \Hc(f_u) \right \}. 
\end{equation*}
Then, it determines the quality level $m_u(t)$ of the requested chunk at time $t$ as:
\begin{equation*} \label{quality-level-decision}
m_u(t) = \mbox{argmin} \left \{Q_{h^*_u(t) u}(t) B_{f_u}(m,t) - \Theta_u(t) D_{f_u}(m,t) \right \},
\end{equation*}
where $B_{f_u}(m,t)$ and $D_{f_u}(m,t)$ are the size in bits and the quality index (could be some subjective measure of video quality; for example SSIM) respectively of chunk $t$ at quality level $m$. 
$\Theta_u(t)$ is a virtual queue introduced to solve the NUM problem. Notice that the streaming of the video file $f_u$ may be handled by different helpers across the streaming session, but each individual chunk is entirely downloaded from a single helper. Notice also that in order to compute the above quantities, each user needs to know only {\em local information} formed by the queue backlog $Q_{hu}(t)$ and the locally computed virtual queue value $\Theta_u(t)$. This scheme is reminiscent of the current adaptive streaming technology for video on demand systems, referred to as DASH (Dynamic Adaptive Streaming over HTTP)\cite{stockhammer2011dynamic,zambelli2009iis}, where the client (user) progressively fetches a video file by downloading successive chunks, and makes adaptive decisions on the quality level based on its current knowledge of the congestion of the underlying server-client connection. Our policy generalizes DASH by allowing the client to dynamically select the least backlogged server, for each chunk. 
 \subsection{Transmission Scheduling}\label{trans-schedul}
 At time slot $t$, the general transmission scheduling consists of maximizing the weighted sum rate of the transmission rates achievable at scheduling slot $t$.
Namely, the network of helpers must solve the Max-Weighted Sum Rate (MWSR) problem:
\begin{align} \label{mwsr-general}
\mbox{maximize} & \;\;\; \sum_{h \in \Hc} \sum_{u \in \Nc(h)} Q_{hu}(t) \mu_{hu}(t) \nonumber \\
\mbox{subject to} & \;\;\; \muv(t) \in \Rc(t)
\end{align}
where $\Rc(t)$ is the region of achievable rates supported by the network at time $t$ and $\mu_{hu}(t)$ is the scheduled rate from helper $h$ to user $u$ in time slot $t$. 
 We particularize the above general MWSR problem to a simple physical layer system.
\subsubsection*{Macro-Diversity}
In this physical layer system, referred to as ``macro-diversity'', the users can decode multiple data streams from multiple helpers if they are scheduled with non-zero rate on the same slot. 
In this case, the rate region $\Rc(t)$ is given by the Cartesian product of the following orthogonal access regions
\begin{equation}  \label{rateconst1}
\sum_{u \in \Nc(h)} \frac{\mu_{hu}(t)}{C_{hu}(t)} \leq 1,  \;\;\;\; \forall~h\in \Hc,
\end{equation}
where $C_{hu}(t)$ is the peak rate from helper $h$ to user $u$ in time slot $t$. 
In the macro-diversity system, the general MWSR problem (\ref{mwsr-general}) decomposes into individual problems, to be solved in a decentralized way at each helper node. The solution is given by each helper $h$ independently choosing the user $u^*_h(t)$ given by 
\begin{equation*}  \label{user-selection}
u^*_h(t) = \mbox{argmax} \left \{ Q_{hu}(t) C_{hu}(t) \; : \; u \in \Nc(h) \right \}, 
\end{equation*}
with rate vector given by $\mu_{h u^*_h(t)}(t) = C_{h u^*_h(t)}(t)$ and $\mu_{hu}(t) = 0$ for all $u \neq u^*_h(t)$. Notice that here, unlike conventional cellular systems, we do not assign a fixed set of users to each helper. In contrast, the helper-user association is dynamic, and results from the transmission scheduling decision. Notice also that, despite the fact that each helper $h$ is allowed to serve its queues with rates $\mu_{hu}(t)$ satisfying (\ref{rateconst1}), the proposed policy allocates the whole $t$-th downlink slot to a single user $u^* \in \Nc(h)$, served at its own peak-rate $C_{hu^*}(t)$.

\subsection{Algorithm Performance}
It can be shown that the time average utility achieved by the proposed policy comes within $O(\frac{1}{V})$ of the utility of a genie-aided T -slot look ahead policy for any arbitrary sample
path with a $O(V)$ tradeoff in time averaged backlog. Thus, the scheme provably achieves optimality of the network utility function under dynamic and arbitrarily changing network conditions; details of the proof can be found in~\cite{bethanabhotla2013joint}.

\subsection{Pre-buffering and Re-buffering Chunks}
The NUM problem formulation (\ref{NUM}) does not take into account the possibility of stall events, i.e., chunks that are not delivered within their playback deadline. This simplification has the advantage of yielding the simple and decentralized scheduling policy described in the previous sections. However, in order to make such policy useful in practice we have to force the system to work in the {\em smooth streaming regime}, i.e., in the regime where the stall events have small probability. This can be done by adaptively determining the pre-buffering time $T_u$ for each user $u$ on the basis of an estimate of the largest delay of queues $\{Q_{hu}: h \in \Nc(u) \}$.

We define the size of the playback buffer $\Psi_t$ as the number of playable chunks in the buffer not yet played. Without loss of generality, assume that the streaming session starts at $t = 1$. Then, $\Psi_t$ is recursively given by the updating equation:\footnote{$1\{\Kc\}$ denotes the indicator function
of a condition or event $\Kc$.}
\begin{align*}
\Psi_t = \max \left \{ \Psi_{t-1}  -  1\{t > T_u\}, 0 \right \} + |a_t|.
\end{align*}
where $|a_t|$ is the number of chunks that are completely downloaded in slot $t$.
Let $A_k$ denote the time slot in which chunk $k$ arrives at the user and let $W_k$ denote the delay with which chunk $k$ is delivered. Note that the longest period during which $\Psi_t$ is not incremented is given by the maximum delay to deliver chunks. Thus, each user $u$ needs to adaptively estimate $W_k$ in order to choose $T_u$. In the proposed method, at each time $t = 1,2,\ldots$, user $u$ calculates the maximum observed delay $E_t$  in a sliding window of size $\Delta$, by letting:
\begin{align}
E_t = \max \{ W_k \; :  ~t-\Delta+1 \leq A_k \leq t \}.
\label{del-window}
\end{align}
Finally, user $u$ starts its playback when $\Psi_t$ crosses the level $\xi E_t$, i.e., $T_u = \min \{ t : ~\Psi_t \geq \xi E_t \}.$
where $\xi$ is a tuning parameter. If a stall event occurs at time $t$, i.e., $\Psi_t = 0$ for $t > T_u$, the algorithm enters a re-buffering phase in which the same algorithm presented above is employed again to determine the new instant $t + T_u + 1$ at which playback is restarted.
\subsection{Extensions}
In~\cite{bethanabhotla2014joint}, we consider extensions and improvements of our work. In Sections \ref{trans-schedul} and \ref{cong-control}, we treated the case of single-antenna base stations and, starting from a network utility maximization (NUM) formulation, we devised a ``push'' scheduling policy, where users place requests to sequential video chunks to possibly different base stations with adaptive video quality, and base stations schedule their downlink transmissions in order to stabilize their transmission queues. In~\cite{bethanabhotla2014joint}, we consider a ``pull'' strategy, where every user maintains a request queue, such that users keep track of the video chunks that are effectively delivered. The pull scheme allows to download the chunks in the playback order without skipping or missing them. In addition, motivated by the recent/forthcoming progress in small cell networks (e.g., in wave-2 of the recent IEEE 802.11ac standard), we extend our dynamic streaming approach to the case of base stations capable of multiuser MIMO downlink, i.e., serving multiple users on the same time-frequency slot by spatial multiplexing. By exploiting the ``channel hardening'' effect of high dimensional MIMO channels, we devise a low complexity user selection scheme to solve the underlying max-weighted rate scheduling~(\ref{mwsr-general}), which can be easily implemented and runs independently at each base station.

\subsection{Preliminary Implementation}
 As observed in \ref{trans-schedul} and \ref{cong-control}, users place their chunk requests from the helpers having the shortest queue pointing at them. Then, transmission scheduling decisions are made by each helper, which maximizes at each scheduling decision time its downlink weighted sum rate where the weights are provided by the queue lengths.  The scheme can be implemented in a decentralized manner, as long as each user knows the lengths of the queues of its serving helpers, and each helper knows the individual downlink rate supported to each served user. Queue lengths and link rates represent rather standard protocol overhead information in any suitable wireless scheduling scheme. We have also implemented a version of such scheme on a testbed formed by Android smartphones and tablets, using standard WiFi MAC/PHY~\cite{kim2013adaptive}.
\section{Performance of D2D Caching Networks}
\label{sec: Performance of D2D Caching Networks}

We now turn to D2D networks, i.e., architectures where the devices themselves act as caches. In contrast to our analysis of femtocaching, we consider here only the transmission of video {\em files} (i.e., no streaming), and also neglect the issue of video rate adaptation (these are topics of ongoing research). In this section, we first outline the principle and intuitive insights. We then discuss the fundamental scaling laws, both for the sum throughput in the cell (disregarding any fairness considerations), and for the tradeoff between throughput and outage. Combining D2D transmission with coding and multicasting is also discussed. 

\subsection{Principle and mathematical model}

As outlined in Sec. III.B, we consider a network where each device can cache a fixed number $M$ video files, and send them - upon request - to other devices nearby. If a device cannot obtain a file through D2D communications, it can obtain it from a macro cellular base station (BS) through conventional cellular transmission. 

Consider a setup in which clustering is used (see Sec. III.B), and assume furthermore deterministic caching. The main performance factor that can be influenced by the system designer is the cluster size; this is regulated through the transmit power (we assume that it is the same for all users in a cell, but can be optimized as a function of user density, library size, and size of the caches). Increasing cluster size increases the probability for finding the desired file in the cluster, while it decreases the frequency reuse. 

There are a number of different criteria for optimizing the system parameters. One obvious candidate is the total network throughput. It is maximized by maximizing the number of active clusters. In \cite{golrezaei2012base}, we showed that for deterministic caching, the expected throughput can be computed as
 \begin{eqnarray} \label{E[y]2}
E \{ T \} & = & \frac{1}{{{r^2}}}\sum\limits_{k = 0}^n \left( 1 - \prod_{i=1}^{k}(1-(P_{CVC}(k)-P_{\rm r}(f_i)))\right) \times \nonumber \\
& & \times \Pr [K = k]. 
\end{eqnarray}
where $P_{CVC}(k)$ is the probability that the requested file is in the Common Virtual Cache (the union of all caches in the cluster), i.e., among the $k$ most popular files. 
$Pr[K=k]$, the probability that there are $k$ users in a cluster, is deterministic for the rectangular grid arrangement, and 
\begin{equation}\label{pr[K=k]}
\Pr [K = k] = \left( \begin{array}{l}
n\\
k
\end{array} \right){(r^2)^k}{(1 - r^2)^{n - k}},
\end{equation}
 for random node placement.

\subsection{Theoretical Scaling Laws analysis}
\label{sec: Theoretical Scaling Laws analysis}

We now turn to  scaling laws, i.e., determine how the capacity scales up as more and more users are introduced into the network. We are dealing with "dense" networks, such that the user density increases, while the area covered by a cell remains the same. As mentioned in Section \ref{sec: File placement for D2D communications}, for the achievable caching scheme, we consider a simple ``decentralized'' random caching strategy, where each user caches $M$ files chosen independently on the library $\Fc$ with probability $P_c^*(f)$ given by (\ref{eq: optimal caching distribution}). 

We furthermore deal again with the "clustered" case, i.e., the network is divided into clusters of equal size $g_c(m)$ 
A system admission control scheme decides whether to serve potential links or ignore them.  The served potential links in the same cluster are scheduled with equal probability (or, equivalently, in round robin), such that all admitted user requests have the same average throughput $\EE[T_u] = \overline{T}_{\min}$ (see \cite{ji2013throughput} for formal definitions.), for all users $u$, where expectation is with respect to
the random user requests, random caching, and the link scheduling policy (which may be randomized or deterministic, as a special case).
To avoid interference between clusters, we use a time-frequency reuse scheme \cite[Ch. 17]{molisch2011wireless} with parameter $K$ 
as shown in Fig.~\ref{fig: Grid_TDMA}.  In particular, we can pick $K = \left(\left\lceil\sqrt{2}(1+\Delta)\right\rceil+1\right)^2$, where $\Delta$ is the interference parameter defined 
in the protocol model. 

In \cite{golrezaei2012wireless_ISIT, golrezaei2012device_Globecom} we established lower and upper bounds for the throughput of D2D communications (this was done under the assumption of random node distribution and caching according to a Zipf distribution). The main conclusion from the scaling law is that for highly concentrated demand distribution, $\gamma_r >1$, the throughput scales linearly with the number of users, or equivalently the per-user throughput remains constant as the user density increases; the number of users in a cluster also stays constant. For heavy-tailed demand distributions, the throughput of the system increases only sub linearly, as the clusters have to become larger (in terms of number of nodes in the cluster), to be able to find requested files within the caches of the cluster members.   

In \cite{ji2013throughput} we sharpened the bounds and extended them to the case of throughput - outage tradeoff.  
Qualitatively (for formal definition see \cite{ji2013throughput}), we say that a user is in outage if the user cannot be served in the D2D network. This can be caused by: (i) the file requested by the user cannot be found in the user's own cluster, 
(ii) that the system admission control decides to ignore the request. We define 
the outage probability $p_o$ as the average fraction of users in outage. At this point, we can define the throughput-outage tradeoff as follows:
\begin{defn} {\bf (Throughput-Outage Tradeoff)}   \label{def: throughput-outage trade-off}
For a given network and request probability mass function $\{ P_r(f) : f \in \Fc\}$,  
an outage-throughput pair $(p,t)$ is {\em achievable} if there exists a cache placement 
scheme and an admission control and  transmission scheduling policy with outage probability
$p_o \leq p$ and minimum per-user average throughput $\overline{T}_{\min} \geq t$. 
The outage-throughput achievable region $\Tc(P_r,n,m)$ is the closure of all achievable outage-throughput pairs $(p,t)$. 
In particular, we let $T^*(p) = \sup \{ t : (p, t) \in \Tc(P_r,n,m) \}$. 
\hfill $\lozenge$
\end{defn}
Notice that $T^*(p)$ is the result of the optimization problem:
\begin{eqnarray} \label{sucaminchia}
\mbox{maximize} & & \overline{T}_{\min} \nonumber \\
\mbox{subject to} & & p_o \leq p, 
\end{eqnarray}
where the maximization is with respect to the cache placement and transmission policies. 
Hence, it is immediate to see that $T^*(p)$ is non-decreasing in $p$.  

The following results are proved in \cite{ji2013throughput} and yield scaling law of the optimal throughput-outage tradeoff
under the clustering transmission scheme defined above. 
%

\begin{figure}[ht]
\centerline{\includegraphics[width=8cm]{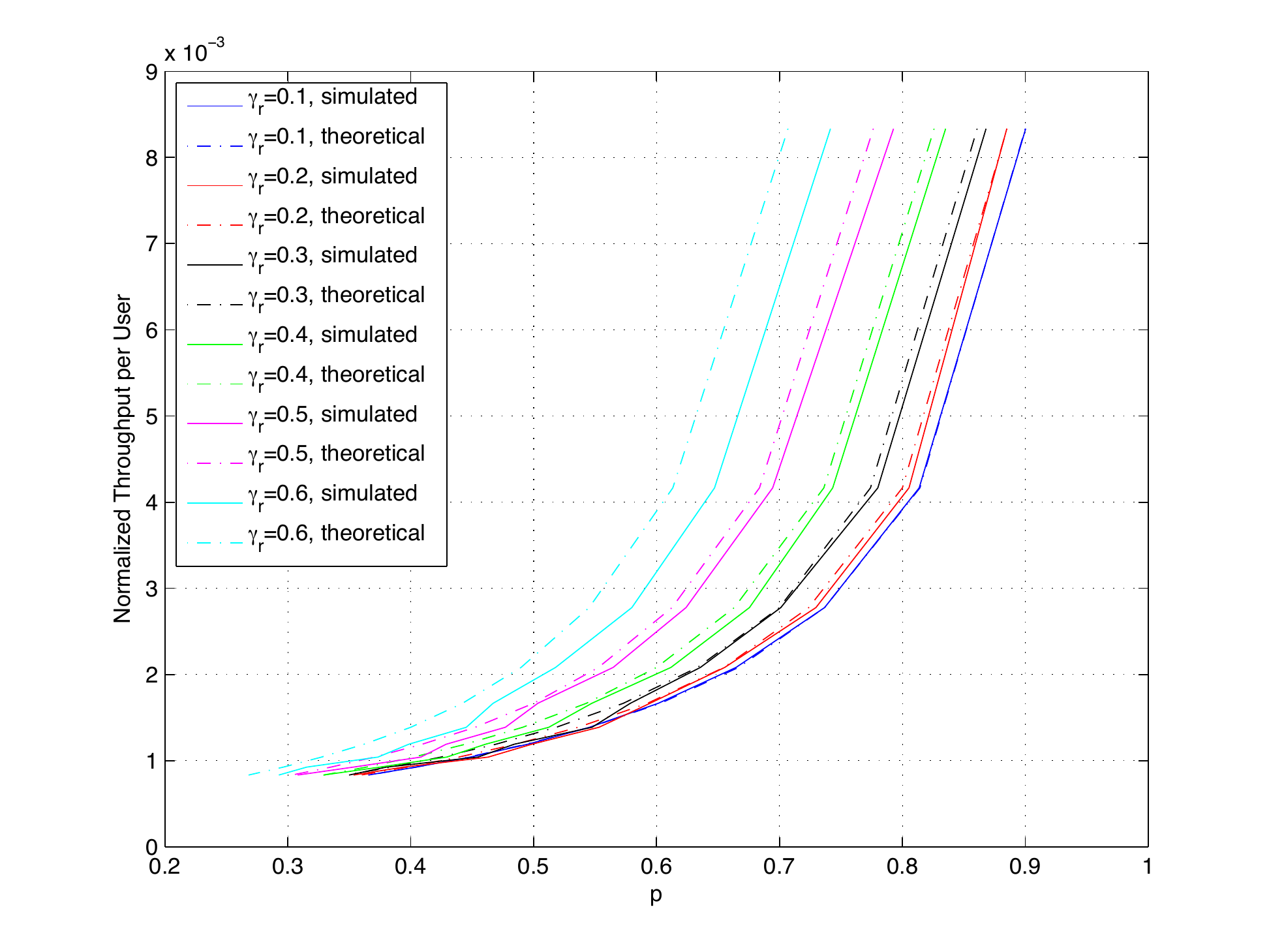}}
\caption{Comparison between the normalized theoretical result and normalized simulated result in terms of the minimum throughput 
per user v.s. outage probability. The throughput is normalized by $C_r$, so that it is independent of the link rate. 
We assume $m=1000$, $n=10000$, and reuse factor $K = 4$.  
The parameter $\gamma_r$ for the Zipf distribution varies from $0.1$ to $0.6$. The theoretical curve  show the plots of 
of the dominating term in (\ref{eq: theorem 4}) divided by $C_r$.}
\label{fig: result_1}
\end{figure}

Although the results of \cite{ji2013throughput} are more general, here we focus on the most relevant regime of the scaling of the file library size with the number of users, referred to as ``small library size'' in \cite{ji2013throughput}. Namely, we assume that $\lim_{n\rightarrow \infty} \frac{m^\alpha}{n} = 0$, where  $\alpha = \frac{1 - \gamma_r}{2 - \gamma_r}$. Since $\gamma_r \in (0,1)$, we have  $\alpha < 1/2$. This means that the library size $m$ can grow even faster than 
quadratically with the number of users $n$. In practice, however, the most interesting case is where $m$ is sublinear with respect to $n$ (see \cite{ji2013throughput} for justifications.). 
Remarkably, any scaling of $m$ versus $n$ slower than $n^{1/\alpha}$ is captured by the following result:  
\begin{theorem} \label{theorem: 4}
Assume $\lim_{n \rightarrow \infty} \frac{m^{\alpha}}{n} = 0$. Then,  the throughput-outage tradeoff achievable by one- hop D2D network 
with random caching and  clustering behaves as:
\begin{align}
\label{eq: theorem 4}
&T^*(p) \geq \notag\\
& \left\{\begin{array}{ll}
\frac{C_r}{K}\frac{M}{\rho_1 m} +  \delta_1(m), & \;\; p = (1-\gamma_r) e^{\gamma_r - \rho_1}, \\
\frac{C_rA}{K} \frac{M}{m (1-p)^{\frac{1}{1-\gamma_r}}} + \delta_2(m),  & \;\; p = 1 - {\gamma_r}^{\gamma_r} \left(\frac{Mg_c}{m}\right)^{1-\gamma_r}, \\
\frac{C_rB}{K } m^{-\alpha} + \delta_3(m),  & \;\; 1 - {\gamma_r}^{\gamma_r} M^{1-\gamma_r} \rho_2^{1-\gamma_r} m^{-\alpha}  \\
& \;\;\;\;\; \leq p \leq 1 - a(\gamma_r)  m^{-\alpha},  \\
\frac{C_rD}{K} m^{-\alpha} + \delta_{4}(m),  & \;\; p \geq 1 - a(\gamma_r) m^{-\alpha},
\end{array}\right.
\end{align}
where $a(\gamma_r)$, $A, B, D$ are some constant depending on $\gamma_r$ and $M$, which can be found in \cite{ji2013throughput},
and where $\rho_1$ and $\rho_2$ are positive parameters satisfying
$\rho_1 \geq \gamma_r$ and $\rho_2 \geq \left(\frac{1-\gamma_r}{\gamma_r^{\gamma_r}M^{1-\gamma_r}}\right)^{\frac{1}{2-\gamma_r}}$. 
The cluster size $g_c$ is any function of $m$ satisfying 
$g_c = \omega\left(m^{\alpha} \right)$ and $g_c \leq \gamma_r m/M$. The functions $\delta_i(m)$, $i = 1,2,3,4$ are vanishing for $m \rightarrow \infty$ with the following orders $\delta_1(m) = o(M/m)$, $\delta_2(m) = o\left(\frac{M}{m(1-p)^{\frac{1}{1-\gamma_r}}} \right)$,
$\delta_3(m)$, $\delta_{4}(m) = o\left(m^{-\alpha}\right)$.\hfill $\square$
\end{theorem}
%
%
The dominant term in (\ref{eq: theorem 4}) can accurately capture the system performance even in the finite-dimensional case shown by simulations in Fig. \ref{fig: result_1}. Further, also in \cite{ji2013throughput}, we can show that the achievable throughput-outage trade-off given by $(\ref{eq: theorem 4})$ is order optimal. When $Mn \geq m$ (the whole library can be cached in the network.), for arbitrarily small outage probability, by using (\ref{eq: theorem 4}), the per user throughput scales as $T^*(p) = \Theta\left(\frac{M}{m}\right)$. This means that the per-user throughput is independent of the number of users (or in other words, the network throughput increases linearly with the number of users, as already indicated above. Furthermore, the throughput grows linearly with $M$. This can be very attractive since, for example, in order to double the throughput, instead of increasing the bandwidth or power, we can just double the (cheap) storage capacity per user.  

Interestingly, our result shown by (\ref{eq: theorem 4}) coincides the achievable throughput by using the subpacketized caching and coded multicasting algorithms in \cite{maddah2012fundamental, Mingyue-D2D-ITW}. However, in realistic channel assumptions, the result is shown in Section \ref{sec: Simulation Results}.

\subsection{Coded caching and multicasting}

\begin{figure}[ht]
\centerline{\includegraphics[width=7cm]{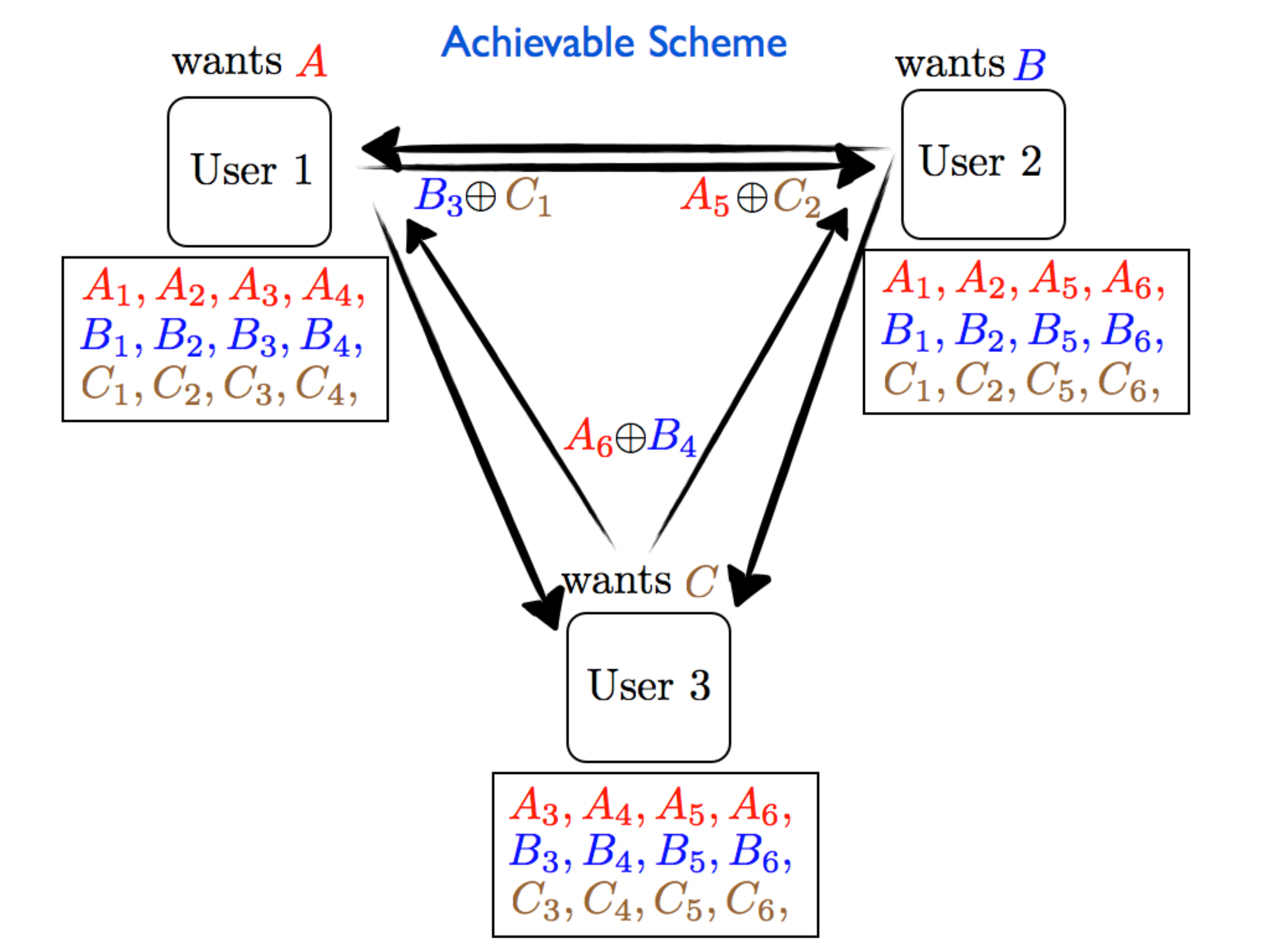}}
\caption{Illustration of the example of $3$ users, $3$ files and $M = 2$, achieving $1/2$ transmissions in term of file. We divide each file into $6$ packets (e.g. $A$ is divided into $A_1, \cdots, A_6$.) We let user 1 requests $A$; user 2 requests $B$ and user 3 requests $C$. The cached packets are shown in the rectangles under each user. For the delivery phase, user $1$ transmits $B_3 \oplus C_1$; user 2 transmits $A_5 \oplus C_2$ and user 3 transmits $A_6 \oplus B_4$. The normalized number of transmissions is $3 \cdot \frac{1}{6} = \frac{1}{2}$, which is also information theoretically optimal for this network \cite{Mingyue-D2D-ITW}. }
\label{abc-user}
\end{figure}

From the previous analysis of the D2D caching network, one important property of the proposed scheme is that in both the caching phase and the delivery phase, an uncoded approach is applied . The gain of the throughput is mainly obtained by spatial reuse (TDMA). At this point, a natural question to ask is whether coded multicasting for D2D transmissions can provide an additional gain, or whether the coding gain and the spatial reuse gain can accumulate. In \cite{Mingyue-D2D-ITW}, we designed a subpacketized caching and a network-coded delivery scheme for the D2D caching networks. The schemes are best to be explained by the example shown in Fig. \ref{abc-user}, where we assume no spatial reuse can be used, or only one transmission per time-frequency slot is allowed but the transmission range can cover the whole network. 
This scheme can be generalized to any $n, m, M$. Without using spatial reuse, for zero outage, the achievable normalized number of transmissions such that every user can successfully decode is $\frac{m}{M}\left(1-\frac{M}{m}\right)$,\footnote{We normalize the number of transmissions by the file size, which is assumed to be same for all the files. } which is surprisingly almost the same as the result shown in \cite{maddah2012fundamental}, where instead of D2D communications, one central server (base station) which has access to all the files multicasts coded packets. In addition, it also has the same scaling law as the throughput by using our previously proposed decentralized caching and uncoded delivery scheme.\footnote{Notice that the reciprocal of the number of transmissions is proportional to the throughput under our protocol model assumption.}
Moreover, it can be shown that there is no further gain when spatial reuse is also exploited. In another word, the gains of spatial reuse and coding cannot accumulate. Intuitively, because if spatial reuse is not allowed, a complicated caching scheme can be designed such that one transmission can be useful for as many users as possible. While if we reduce transmission range and perform our scheme in one cluster as shown in Fig. \ref{fig: Grid_TDMA}, then the number of users benefitted by one transmission is reduced but the D2D transmissions can operate simultaneously at a higher rate. Moreover, the complexity of caching subpactization and coding can also be reduced.  Hence, the benefit of coding depends on the actual physical layer throughput (bits/s/Hz) and the caching/coding complexity rather than throughput scaling laws. 

\subsection{Simulation Results}
\label{sec: Simulation Results}

To see the difference between the performance of the proposed D2D caching network and the state-of-the-art schemes for video streaming, we need to consider the realistic propagation and interference channel mode instead of the protocol model. One reason is that as mentioned in Section \ref{sec: Theoretical Scaling Laws analysis}, for small outage probability, the throughput of the proposed D2D scheme has the same scaling laws as the coded multicasting scheme in \cite{maddah2012fundamental}. The state-of-the-art schemes that will be compared with are conventional unicasting, harmonic broadcasting and coded multicasting, whose details can be found in \cite{JiCaireMolisch2013}. In the following, for practice considerations, the proposed uncoded D2D scheme discussed in Section \ref{sec: Theoretical Scaling Laws analysis} is used for simulations. 

\begin{figure}[ht]
\centerline{\includegraphics[width=8cm]{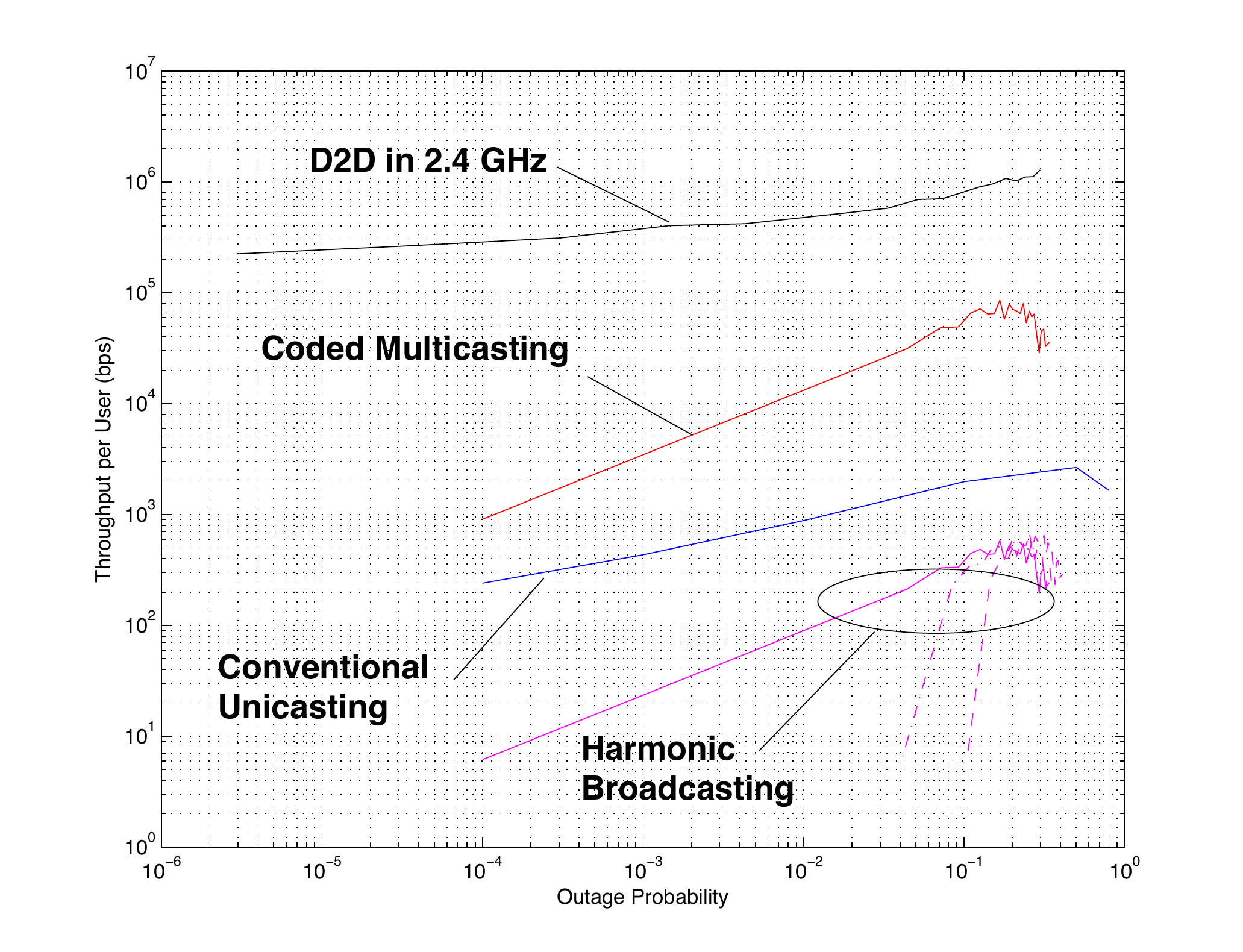}}
\caption{Simulation results for the throughput-outage tradeoff for different schemes under the realistic indoor/outdoor propagation environment 
(for details, see \cite{JiCaireMolisch2013}). For harmonic broadcasting with only the most $m'$ popular files, solid line: $m'=300$; dash-dot line: $m'=280$; dash line: $m'=250$. We have $n = 10000$, $m = 300$, $M = 20$ and $\gamma_r = 0.4$.}
\label{fig: result_2_new}
\end{figure}

For simulations, we considered a network of size $600 {\rm m} \times 600 {\rm m}$, 
where we relax the grid structure of the users' distribution and let $n = 10000$ users distributed uniformly. 
The file library has size $m = 300$ (e.g., 300 popular movies and TV shows to be refreshed on a daily basis at off-peak times 
by the cellular network).  The storage capacity in each user is $M = 20$ and the parameter for the Zipf distribution is $\gamma_r = 0.4$ \cite{breslau1999web}. 
We considered a regular patterns of buildings of size $50{\rm m} \times 50{\rm m}$, separated by streets
of widths $10$m \cite{JiCaireMolisch2013}, with indoor, outdoor, indoor-to-outdoor and outdoor-to-indoor pathloss and shadowing
models taken from \cite{winner2007d1}, assuming that D2D links operate at $2.4$GHz (WiFi-direct). We assumed a channel bandwidth of $20$ MHz 
in order to provide throughput in bit/s.  All the details of the simulation parameters, including the pathloss and shadowing models, 
can be found in \cite{JiCaireMolisch2013}.
The simulation results of the throughput-outage tradeoff for different schemes are given in 
Fig.~\ref{fig: result_2_new}. We observe that in this realistic propagation scenario the D2D single-hop caching network 
can provide  both large throughput, sufficient for streaming video at standard definition quality, and low outage probability. 
Also, the D2D caching scheme significantly outperforms the other schemes in the regime of low outage probability.  
This performance gain is particularly impressive with respect to conventional unicasting and harmonic broadcasting 
from the base station, which are representative of the current technology. We also note the distinct performance advantages compared to coded multicasting - despite the fact that
the two schemes have the same scaling laws. The main reason for this development is that the capacity of multicasting is limited by the "weakest link" between BS and the various MSs, while for the D2D transmission scheme, short distance transmission (which usually has high SNR, shallow fading, and thus high capacity) determine the overall performance. 

It is also worthwhile to notice that the scheduling scheme used in the simulations is based on the clustering structure and the interference avoidance (TDMA) discussed in Section \ref{sec: Theoretical Scaling Laws analysis} without using any advanced interference management scheme such as FlashLinQ \cite{wu2010flashlinq} and ITLinQ \cite{naderializadeh2013itlinq}, which may provide an even higher gain in terms of throughput for the D2D caching networks.

\section{Conclusions}

As user demand for video data continues to increase sharply in cellular networks, new approaches are needed to dramatically expand network capacity.  This paper has provided an overview of an approach explored by the University of Southern California as part of the industry-sponsored research program, Video Aware Wireless Networks (VAWN).  The approach exploits a key feature of wireless video, namely the high degree of (asynchronous) content reuse across users.  To exploit this feature, we propose replacing expensive backhaul infrastructure with inexpensive caching capabilities.  This can be realized in two ways:  the use of Femto-Caching or dedicated helper nodes that cache popular files and serve nearby user requests, and the use of user devices themselves to cache and exchange files using device-to-device (D2D) communications. Simulations with realistic settings show that even for relatively low-density deployment of helper stations, throughput can be increased by a factor five. D2D networks allow in many situations a throughput increase that is linear with the number of users (thus making the per-user throughput independent of the number of users). Simulations in realistic propagation channels, storage capacity settings, video popularity distributions, and user densities show that (for constant outage), the throughput can be two orders of magnitude or more higher than the state-of-the-art multicast systems. 
 
A key issue in our caching approach is that of file placement.  In the helper node approach, we show that the problem of minimizing average file downloading time in the uncoded placement case (video-encoded files are cached directly on help nodes) is NP-complete, but can be reformulated and is solvable as a monotone submodular function over matroid constraints.  For the coded case (coded chunks of files are placed on different helper stations), optimum cache placement can be formulated and is solvable as a convex optimization problem.  Also for the D2D approach, the question of which files to cache is key.   Two approaches are deterministic caching in which a BS instructs devices which files to cache (i.e., the most popular and in a disjoint manner), and random caching in which each device randomly caches a set of files according to a probability mass function. It is remarkable that the simple random caching is not only optimum from a scaling-law point of view, but also in numerical simulations provides throughputs that are close to the deterministic caching (which is ideal but difficult to realize for time-varying topologies). 
 
An important area of future work is that of predicting user requests.  The effectiveness of caching schemes depends not only on the degree of content reuse, but on our ability to understand and predict request behavior across clusters of users. Furthermore, the approach is predicated on a "time-scale decomposition", namely that request distributions change much more slowly (over days or weeks) than the time it takes to stream a video (minutes to a couple of hours). For femto-caching, it is noteworthy that the type of users (and thus the requests) within range of a helper station might change over the course of a day; more research on how such spatio-temporal aspects can be predicted and accommodated is required. Similarly, the impact of social networks on user preferences could be exploited.  

In the D2D sphere, research on new approaches for incentivizing users to participate in cooperative caching schemes is needed.  Both helper node and D2D caching schemes would benefit from research into multi-hop cache retrieval schemes and PHY schemes that better exploit advances in wireless communication technology (e.g., multiuser MIMO).  In the D2D area, we are/will be investigating how to optimize neighbor discovery, estimating channel conditions and then using the information to make scheduling optimizations, and transmission schemes closely tuned to existing communications standards like WiFi Direct.

\pagenumbering{arabic}
\def\thepage{D-\arabic{page}}
\bibliographystyle{IEEEtran}
\bibliography{dilip-ref,references_Dilip,references_Intel,references_Mol,references}

\end{document}